# 2D Hydrogenated Graphene-like Borophene as a High Capacity Anode Material for Improved Li/Na Ion Batteries: A First Principles Study

Meysam Makaremi,[1] Bohayra Mortazavi,[2] and Chandra Veer Singh*[,1,3]

[1]Department of Materials Science and Engineering, University of Toronto, 184 College Street, Suite 140, Toronto, ON M5S 3E4, Canada.
[2]Institute of Structural Mechanics, Bauhaus-Universität Weimar, Marienstr. 15, D-99423 Weimar, Germany.
[3]Department of Mechanical and Industrial Engineering, University of Toronto, 5 King's College Road, Toronto M5S 3G8, Canada.

**Abstract**

Fast-growing electronics industry and future energy storage needs have encouraged the design of rechargeable batteries with higher storage capacities, and longer life times. In this regard, two-dimensional (2D) materials, specifically boron and carbon nanosheets, have garnered enthusiasm due to their fascinating electronic, optical, mechanical and chemical properties. Recently, a hydrogen boride (HB) nanosheet was successfully fabricated showing remarkable stability and superior physical properties. Motivated by this experimental study, we used first principle electronic structure calculations to study the feasibility of this nanosheet to serve as an anode material for Li/Na/Ca/Mg/Al ion batteries. Most active adsorption sites for single adatoms were evaluated and next adatoms were gradually inserted into the anode surface accordingly. The charge transfer, electronic density of sates, storage capacity, structural stability, open-circuit potential and diffusion energy barriers were explored. Our theoretical study predicts that HB shows outstanding electrode properties for Li and Na ion batteries. The intercalation of both Li and Na adatoms into the HB monolayer can lead to a high identical storage capacity of 1133.8 mAh/g which is promising compared to the capacities of the traditional anode materials; such as graphite (372 mAh/g) and $TiO_2$ (200 mAh/g), and other 2D materials; such as germanene (369 mAh/g), stanene (226 mAh/g), and phosphorene (432.8 mAh/g) nanosheets. These results may open a new horizon for the design of rechargeable batteries with higher storage capacitates.

## 1. Introduction

Rechargeable metal-ion batteries play critical roles for a variety of advanced technologies such as mobile-electronics, communication devices and electric vehicles.[1–4] In this regard, the use of advanced materials with higher charge capacities and faster ion diffusion rates are extremely in demand to replace common anode and cathode materials. Among various candidates, 2D materials and their hybrid structures are among the most promising solutions owing to their large surface area and superior electronic, thermal and mechanical properties.[5–11] Previous theoretical studies have confirmed that 2D materials can yield remarkably high charge capacities[12–15] and ultralow diffusion energy barriers[16–21] for Li and Na-ion batteries.

Graphene's outstanding physical, chemical and electronic properties[22–24] have led to the introduction of the diverse family of two-dimensional (2D) materials.[25] A few of which are single elemental nanosheets; such as phosphorene, silicene, germanene and stanene; although, they cannot forms versatile bonding like $sp^3$, $sp^2$ and sp like graphene[26], these structures present limited bond length flexibility, and their equilibrated single-layers exist in out-of-plane buckled configurations. On the other side, boron, the neighboring element of carbon in the periodic table, can similarly form diverse bonding and it involves diverse forms, from zero-dimensional to three-dimensional crystals.[26–29]

During the last couple of years, 2D boron sheets so called borophene with the buckled[30] and flat[31] atomic configurations have been successfully fabricated using the epitaxy growth of boron atoms on the silver metallic surface. Followed by these experimental reports, numerous theoretical studies have been achieved to explore the application of 2D boron based nanomembranes for various applications such as hydrogen storage[32,33], rechargeable metal-ion batteries[34–40], superconducting, magnetic, electronic, and chemical devices[39,41–45], and mechanically robust components.[46–48] Most recently, experimental realization and characterization of 2D hydrogen boride sheets with an empirical formula of $H_1B_1$ was successfully achieved by exfoliation and complete ion-exchange between protons and magnesium cations in magnesium diboride.[49] Worthy to note that the existence and stability of such a 2D structure has already been theoretically proven.[50]

In this regard, pristine and functionalized borophene films have been also theoretically realized to yield remarkably high charge capacities, good electronic conductivity and low energy diffusion barriers, which are all highly desirable for the application as anode materials in the rechargeable battery industry.[32–36,51] 2D borophene structures were synthesized by growing on a substrate, and develop of efficient transferring methods for lifting 2D nanosheets from the metallic substrate in order to reach isolated nanomembranes has been a significant challenge to date.[52] Moreover, for the application in rechargeable batteries, borophene films have to be producible in large content and low-cost.[36] Interestingly, a recently fabricated hydrogenated borophene structure, hydrogen boride[49], was in multi-layer form and more importantly the nanomembranes were not grown on a substrate, which accordingly can serve as promising signs toward their real application in commercial ion batteries.

The successful synthesis of hydrogen boride (HB) raises an important question, whether this new nanomembrane can serve as a promising anode material in rechargeable batteries or not. To address this important question, we conducted extensive density functional theory (DFT) simulations to explore the possible application of the HB monolayer for Na, Ca, Al, Mg, and Li-ion batteries. Our first-principles analysis of charge capacity, adsorption energies, open-circuit voltage profiles, dynamical stabilities, electronic structures and diffusion energy barriers confirm the promising candidacy of hydrogen boride for Na and Li-ions storage.

## 2. Computational Details

Implemented in the Vienna Ab-initio Simulation Package (VASP)[53] framework, Spin polarized Perdew–Burke–Ernzerhof (PBE)[54] density functional theory (DFT) calculations were used through generalized gradient approximation (GGA) and projector augmented-wave (PAW) potentials[55]. A large kinetic energy cutoff of 500 eV. Since the GGA underestimates the binding energies[56,57], a Grimme dispersion correction approach, DFT-D2[58] were selected to accurately calculate energies comparable with experimental cohesive energies as well as binding energies. For electronic self-consistency and ionic relaxation convergence, criteria of 1x10$^{-6}$ eV and 1x10$^{-3}$ eV/Å were considered. And to integrate the Brillouin zone, a Monkhorst-Pack mesh of 15x15x1 and the tetrahedron scheme with Blöchl corrections were applied.

The binding energy were calculated as,

$$E_{Bind} = \frac{(E_{HB+M} - E_{HB} - n \times E_M)}{n}, \qquad (1)$$

where $E_{HB+M}$, $E_{HB}$, $E_M$, and n are the total interaction energy, the energy of the pristine nanosheet, the energy of a single adatom in the gas phase, and the number of intercalating foreign atoms; accordingly. More negative values of $E_{Bind}$ indicates the stronger binding between the surface and adatoms.

The Bader analysis technique[59] were employed to probe charge gain and loss. The system charge difference can be determined by,

$$\Delta\rho = \rho_{HB+M} - \rho_{HB} - \rho_M, \qquad (2)$$

where, $\rho_{HB+M}$, $\rho_{HB}$, and $\rho_M$ define the electron densities of HB+M, HB, and metallic foreign atom, respectively.

Open-circuit voltage $(V)^{60}$ is a critical parameter indicating the performance of an anode electrode, and it can be calculated by using the following equation,

$$V \approx \frac{\left((x_2-x_1)E_{M_B} - (E_{M_{x_2}HB} - E_{M_{x_1}HB})\right)}{(x_2-x_1)e}, \qquad (3)$$

here, $x_1$, $x_2$, and x ($x_1 \leq x \leq x_2$), respectively, involve the initial, final and average coverage ratios. e is the electron charge; and $E_{M_{x_1}HB}$, $E_{M_{x_2}HB}$, and $E_{M_B}$ are the interaction energy for the coverage ratio of $x_1$ and $x_2$, and the metallic bulk energy, respectively.

As illustrated in Figure 1, the HB unitcell consists of 8 (4 hydrogen and 4 boron) atoms and involves an orthorhombic structure with lattice constants a = 3.02 Å and b = 5.29 Å. More details about the HB structure are listed in Tables S1 and S2. For all simulations, we used a supercell composed of 4 × 2 unitcells including 32 H and 32 B atoms, and a vacuum space of 20 Å in the z direction. For each type of adatom, four possible sites were analyzed to find the most favorable adsorption point (see Fig. 1b).

To calculate the battery storage capacity, the adatoms were gradually added to the nanosheet at predefined sites randomly and uniformly until the maximal surface coverage with the highest possible capacity were achieved. Storage capacity is calculated by using Faraday formula, as follows:

$$q = 1000 \, F \, z \frac{n_{max}}{M_{HB}}, \qquad (4)$$

here F, z, $n_{max}$, and $M_{HB}$ are the Faraday constant, atomic valence number, maximal adatom surface coverage, and the molecular mass of the HB single layer.

Thermal stability of relaxed structures was analyzed by performing ab-initio molecular dynamics (AIMD) simulations in canonical ensemble (NVT) at 300 K with a time step of 1 fs for 10000 steps.

## 3. Results and Discussion

As shown in Fig. 1b, the binding energy of the insertion of five types of adatoms consisting of Li, Na, Mg, Ca, and Al at four possible sites of the HB monolayer were calculated to evaluate the ability of HB nanosheet as an anode electrode for alkali and alkali-earth ion batteries. The calculations predict that all of the foreign atoms prefer to be adsorbed at the hexagonal hollow site (Site 4 in Fig. 1b) resulting in adsorption energies of -2.32, -1.27, -0.23, -1.46, and -2.17 eV for Li, Na, Mg, Ca, and Al; respectively. Also, the differential charge density plots of the relaxed structures of the most stable configurations of the monolayer interacting with two types of adatoms, Li and Na, are shown in Figure 2. It suggests that the HB surface strongly interacts with both alkali elements and accept electron charge densities from them. The binding energy calculations predict pretty intense binding between the HB surface and most of the adatoms (except Mg); however, since the initial voltage for the intercalation of Mg, Ca, and Al into the surface is negative (-1.56, -0.60, and -1.57 V for Mg, Ca, and Al; respectively.), the monolayer cannot be applied for these types of rechargeable batteries. Therefore, in this study we specifically focused on anode properties of the HB nanosheet for Li/Na ion storage applications. We added different number of Li/Na adatoms to the HB monolayer (see Figure S1 and S2), and investigated its response as an anode material.

It is worthwhile nothing that the electronic conductivity of an anode electrode is a critical parameter influencing the performance and controlling its internal electronic resistance. Since during charging and discharging, a battery generate the ohmic heat, a proper anode need to be metallic and keep this property during charge/discharge cycles.[35] The density of states (DOS) diagrams for HB monolayers covered with various number of adatoms are presented in Figure 3 suggesting that the pure HB system and all of the systems involving both HB and foreign atoms contain the zero-energy gap and present the conducting behavior.

Fig. 4a shows the HB binding energy as a function of adatom concentration. Due to the smaller size and higher activity of the Li element compared to that of the Na type, as presented in Figs. 4b and 4c the former group of atoms tend to migrate closer to the surface. Therefore, the insertion of Li adatoms leads to more negative energies compared to the binding of the latter counterparts. It should be noted that the energy remains almost constant (~ -1.3 eV) during the whole Li adsorption spectrum, while it increases with respect to the Na intercalation for $x > 0.5$.

The open-circuit voltage is a key aspect showing the performance and the capacity of an anode electrode. The negative values of the potential difference suggest that foreign adatoms prefer to form metallic clusters instead of adsorption to the electrode. It should be noted that the favored potential range for anode materials involves a spectrum of 0.1-1 V.[18] Fig. 5a depicts the potential as a function of storage capacity, and illustrates that the HB anode contains average voltages of 0.65 vs Li/Li$^+$ and 0.03 V vs Na/Na$^+$ for the Li and Na insertion, respectively, which remain constant during the whole adsorption process until reaching the optimal capacity point of $x = 0.5$. For comparison, the potential range for flat borophene[36], borophane[35], and TiO$_2$[61] anodes reported to be 0.5-1.8, 0.03-0.6, and 1.5-1.8 V; respectively.

It is worthwhile to note that the HB anode presents an outstanding identical capacity of 1133.8 mAh/g for both of the Li and Na ion storage batteries at $x = 0.5$. Before the optimal capacity ($x <= 0.5$) the intercalation behavior of Li adatoms is consistent with that of Na ones, leading to the creation of a single layer of adatoms in each side of the monolayer (see Figs. 4b and 4c). However, after reaching the optimal point, as can be seen in Fig. 5a the binding energies are not similar for the two types of adatoms resulting in two different adsorption scenarios. After the optimal concentration (x > 0.5), highly active Li atoms destroy the monolayer structure by attacking the HB bonds (see Fig. 5b). Whereas, Na atoms prefer to limit the anode capacity by creating the second layer of adatoms instead of destroying the monolayer structure (see Fig. 5c), and preventing additional electronic charge transfer (see Figure S3). After reaching the optimal capacity, the stability of each structure was evaluated by AIMD simulations at 300 K. The variations of the temperature as a function of the simulation time is illustrated in Figure S4. Our AIMD simulations confirm that the HB monolayer remains intact upon the adatoms adsorption and represents thermal stability for the anode material usage.

For comparison it should be noted that the theoretical charge capacities of commercial graphite[62] and $TiO_2$[63] structures for the Li ion storage were reported to be 372 and 200 mAh/g, respectively. 2D materials including germanene[64], stanene[64], phosphorene[65] nanosheets illustrate storage capacities of 369, 226, and 432.79 mAh/g, accordingly. Also, studied in the last couple of years, boron 2D structures consisting of buckled borophene[66], flat borophene[31], and borophane[35] were predicted to include capacities of 1720, 1980, and 504 mAh/g, respectively.

It is worthwhile noting that among all 2D materials considered in the literature for the anodic application, only a few of them including flat and buckled borophene present higher storage capacities compared to hydrogen boride. Nevertheless, one should not forget that, these high capacity materials have been synthesized by growing on a substrate and their isolated layers have not been fabricated yet. Whereas, the hydrogen boride nanosheet has been successfully fabricated and isolated by a cation-exchange exfoliation method, which presents stability required for the metal-ion battery application and further for the mass production.

Diffusion of adatoms through the anode material is an essential characteristic indicating the charge/discharge ability of the material. We used the Nudged-elastic band (NEB) method to find the minimum energy paths. The result for the diffusion of Li/Na adatoms on the monolayer through various possible pathways is depicted in Figure 6. For both adatoms, the diffusion via Path 2 (over the B-B band) leads to the lowest energy barriers involving 0.86 and 0.32 eV for Li and Na; respectively. The lower barrier for the Na adatom might be described by weaker interaction between the Na atom with the HB surface (-1.27 eV) compared to that of the small Li element (-2.32 eV), which causes an easier and faster transfer rate for this adatom. The Li and Na energy barriers for the HB nanosheet are close to the data for flat borophene including 0.69 and 0.34 eV reported by an earlier theoretical work[36]. It should be mentioned that the diffusion barrier for Li adatom transfer on other 2D monolayers such as graphene[67], CEY graphene[68], silicene[69], $MoS_2$[70], and $Ti_3C_2$ MXene[71] consists of 0.32, 0.58, 0.23, 0.25, and 0.7 eV; respectively. Moreover, the energy barrier for commercial anode materials based on $TiO_2$ ranges from 0.35 to 0.65 eV[72,73], and for graphite[74] it is around 0.4 eV.

In addition, Figs. 6b and 6c show that the movement of single adatoms along other paths is more energetically demanding. The barriers for the Li and Na movement over the hydrogen atom (along Path 2) are 1.42 and 0.66 eV, and through the hexagonal hole (along Path 3) consist of 10.41 (not shown in Fig. 6c) and 2.21 eV; respectively. The result suggests that diffusion of Na along the latter path might be impossible due to the larger radius of this element compared to that of Li.

## 4. Conclusions

Extensive spin polarized DFT-D2 simulations were carried out to investigate the performance of a newly fabricated hydrogenated graphene-like borophene, hydrogen boride, nanosheet to serve as an anode material for Li/Na/Mg/Ca/Al ion storage applications. Different calculations such as adsorption energy, average potential, diffusion barrier, density of states and Bader charge analyses were conducted. It is found that while the insertion of Mg, Ca, and Al into the nanosheet may not be feasible, the binding of Li and Na can result in superior anode electrodes. Our calculations suggest that the HB monolayer interacting with Li and Na adatoms show high electronic conductivity, drastic structural stability, low charging voltage, and versatile storage capacity.

Diffusion of Li and Na adatoms upon the monolayer includes barrier energies of 0.86 and 0.32 eV, accordingly. The Li and Na insertions result in open-circuit potentials of 0.65 vs Li/Li$^+$ and 0.03 V vs Na/Na$^+$. The intercalation of both types of adatoms into HB leads to an identical capacity of 1133.8 mAh/g resulting in one of the highest storage capacities ever calculated for 2D nanosheets. Only a few 2D nanostructures were theoretically predicted to illustrate similar or higher capacities; although, since these structures were not physically isolated, yet they cannot be used as anode materials. The result of this study is promising and we hope it will shed light on the development of the new generation of Li/Na ion batteries for modern technologies including portable electronics, communication applications, aerospace devices and electric vehicles.

AUTHOR INFORMATION

**Corresponding Author**

* chandraveer.singh@utoronto.ca

ACKNOWLEDGMENT

MM and CVS gratefully acknowledge their financial support in parts by Natural Sciences and Engineering Council of Canada (NSERC), University of Toronto, Connaught Global Challenge Award, and Hart Professorship. The computations were carried out through Compute Canada facilities, particularly SciNet and Calcul-Quebec. SciNet is funded by the Canada Foundation for Innovation, NSERC, the Government of Ontario, Fed Dev Ontario, and the University of Toronto, and we gratefully acknowledge the continued support of these supercomputing facilities. BM greatly acknowledges the financial support by European Research Council for COMBAT project (Grant number 615132).

# References

(1) Sun, Y.; Liu, N.; Cui, Y. Promises and Challenges of Nanomaterials for Lithium-Based Rechargeable Batteries. *Nature Energy*. 2016.
(2) Xu, J.; Dou, Y.; Wei, Z.; Ma, J.; Deng, Y.; Li, Y.; Liu, H.; Dou, S. Recent Progress in Graphite Intercalation Compounds for Rechargeable Metal (Li, Na, K, Al)-Ion Batteries. *Advanced Science*. 2017.
(3) Liu, T.; Ding, J.; Su, Z.; Wei, G. Porous Two-Dimensional Materials for Energy Applications: Innovations and Challenges. *Materials Today Energy*. 2017, pp 79–95.
(4) Shi, Z. T.; Kang, W.; Xu, J.; Sun, Y. W.; Jiang, M.; Ng, T. W.; Xue, H. T.; Yu, D. Y. W.; Zhang, W.; Lee, C. S. Hierarchical Nanotubes Assembled from $MoS_2$-Carbon Monolayer Sandwiched Superstructure Nanosheets for High-Performance Sodium Ion Batteries. *Nano Energy* **2016**, *22*, 27–37.
(5) Thackeray, M. M.; Wolverton, C.; Isaacs, E. D. Electrical Energy Storage for Transportation—approaching the Limits Of, and Going Beyond, Lithium-Ion Batteries. *Energy Environ. Sci.* **2012**, *5* (7), 7854.
(6) Novoselov, K. S.; Mishchenko, A.; Carvalho, A.; Castro Neto, A. H. 2D Materials and van Der Waals Heterostructures. *Science (80-. )*. **2016**, *353* (6298), aac9439.
(7) Wang, X.; Xia, F. Van Der Waals Heterostructures: Stacked 2D Materials Shed Light. *Nature Materials*. 2015, pp 264–265.
(8) Geim, A. K.; Grigorieva, I. V. Van Der Waals Heterostructures. *Nature*. 2013, pp 419–425.
(9) Xu, M.; Liang, T.; Shi, M.; Chen, H. Graphene-like Two-Dimensional Materials. *Chemical Reviews*. 2013, pp 3766–3798.
(10) Tang, J.; Chen, D.; Yao, Q.; Xie, J.; Yang, J. Recent Advances in Noble Metal-Based Nanocomposites for Electrochemical Reactions. *Materials Today Energy*. 2017, pp 115–127.
(11) Zhu, Y.; Chen, G.; Zhong, Y.; Zhou, W.; Liu, M.; Shao, Z. An Extremely Active and Durable $Mo_2C$/graphene-like Carbon Based Electrocatalyst for Hydrogen Evolution Reaction. *Mater. Today Energy* **2017**, *6*, 230–237.
(12) Hwang, H.; Kim, H.; Cho, J. $MoS_2$ Nanoplates Consisting of Disordered Graphene-like Layers for High Rate Lithium Battery Anode Materials. *Nano Lett.* **2011**, *11* (11), 4826–4830.
(13) Eftekhari, A. Molybdenum Diselenide ($MoSe_2$) for Energy Storage, Catalysis, and Optoelectronics. *Appl. Mater. Today* **2017**, *8*, 1–17.
(14) Hu, J.; Xu, B.; Yang, S. A.; Guan, S.; Ouyang, C.; Yao, Y. 2D Electrides as Promising Anode Materials for Na-Ion Batteries from First-Principles Study. *ACS Appl. Mater. Interfaces* **2015**, *7* (43), 24016–24022.
(15) Zhou, Q.; Wu, M.; Zhang, M.; Xu, G.; Yao, B.; Li, C.; Shi, G. Graphene-Based Electrochemical Capacitors with Integrated High-Performance. *Mater. Today Energy* **2017**, *6*, 181–188.
(16) Samad, A.; Noor-A-Alam, M.; Shin, Y.-H. First Principles Study of a $SnS_2$/graphene Heterostructure: A Promising Anode Material for Rechargeable Na Ion Batteries. *J. Mater. Chem. A* **2016**, *4* (37), 14316–14323.
(17) Samad, A.; Shafique, A.; Kim, H. J.; Shin, Y.-H. Superionic and Electronic Conductivity in Monolayer $W_2C$: Ab Initio Predictions. *J. Mater. Chem. A* **2017**, *5* (22), 11094–11099.

(18) Cak[i without dot]r, D.; Sevik, C.; Gulseren, O.; Peeters, F. M. Mo2C as a High Capacity Anode Material: A First-Principles Study. *J. Mater. Chem. A* **2016**, *4* (16), 6029–6035.
(19) Liu, Y.; Peng, X. Recent Advances of Supercapacitors Based on Two-Dimensional Materials. *Appl. Mater. Today* **2017**, *8*, 104–115.
(20) Mortazavi, B.; Shahrokhi, M.; Makaremi, M.; Rabczuk, T. Theoretical Realization of Mo$_2$P; a Novel Stable 2D Material with Superionic Conductivity and Attractive Optical Properties. *Appl. Mater. Today* **2017**, *9*.
(21) Sengupta, A.; Frauenheim, T. Lithium and Sodium Adsorption Properties of Monolayer Antimonene. *Mater. Today Energy* **2017**, *5*, 347–354.
(22) Novoselov, K. S.; Geim, A. K.; Morozov, S. V; Jiang, D.; Zhang, Y.; Dubonos, S. V; Grigorieva, I. V; Firsov, A. A. Electric Field Effect in Atomically Thin Carbon Films. *Science* **2004**, *306* (5696), 666–669.
(23) Geim, A. K.; Novoselov, K. S. The Rise of Graphene. *Nat. Mater.* **2007**, *6* (3), 183–191.
(24) Li, W.; Zhang, Z.; Tang, Y.; Bian, H.; Ng, T.-W.; Zhang, W.; Lee, C.-S. Graphene-Nanowall-Decorated Carbon Felt with Excellent Electrochemical Activity Toward VO 2 + /VO 2+ Couple for All Vanadium Redox Flow Battery. *Adv. Sci.* **2015**, *3*, 1500276 1-7.
(25) Roome, N. J.; Carey, J. D. Beyond Graphene: Stable Elemental Monolayers of Silicene and Germanene. *ACS Appl. Mater. Interfaces* **2014**, *6* (10), 7743–7750.
(26) Zhang, Z.; Penev, E. S.; Yakobson, B. I. Two-Dimensional Boron: Structures, Properties and Applications. *Chem. Soc. Rev.* **2017**.
(27) Yang, Y.; Zhang, Z.; Penev, E. S.; Yakobson, B. I. B40 Cluster Stability, Reactivity, and Its Planar Structural Precursor. *Nanoscale* **2017**, *9* (5), 1805–1810.
(28) Kondo, T. Recent Progress in Boron Nanomaterials. *Sci. Technol. Adv. Mater.* **2017**, *18* (1), 780–804.
(29) Zhang, Y.; Wu, Z. F.; Gao, P. F.; Zhang, S. L.; Wen, Y. H. Could Borophene Be Used as a Promising Anode Material for High-Performance Lithium Ion Battery? *ACS Appl. Mater. Interfaces* **2016**, *8* (34), 22175–22181.
(30) Mannix, A. J.; Zhou, X.-F.; Kiraly, B.; Wood, J. D.; Alducin, D.; Myers, B. D.; Liu, X.; Fisher, B. L.; Santiago, U.; Guest, J. R.; Yacaman, M. J.; Ponce, A.; Oganov, A. R.; Hersam, M. C.; Guisinger, N. P. Synthesis of Borophenes: Anisotropic, Two-Dimensional Boron Polymorphs. *Science (80-. ).* **2015**, *350* (6267), 1513–1516.
(31) Feng, B.; Zhang, J.; Zhong, Q.; Li, W.; Li, S.; Li, H.; Cheng, P.; Meng, S.; Chen, L.; Wu, K. Experimental Realization of Two-Dimensional Boron Sheets. *Nat. Chem.* **2016**, *8*, 563–568.
(32) Li, L.; Zhang, H.; Cheng, X. The High Hydrogen Storage Capacities of Li-Decorated Borophene. *Comput. Mater. Sci.* **2017**, *137* (Supplement C), 119–124.
(33) Rao, D.; Zhang, L.; Meng, Z.; Zhang, X.; Wang, Y.; Qiao, G.; Shen, X.; Xia, H.; Liu, J.; Lu, R. Ultrahigh Energy Storage and Ultrafast Ion Diffusion in Borophene-Based Anodes for Rechargeable Metal Ion Batteries. *J. Mater. Chem. A* **2017**, *5* (5), 2328–2338.
(34) Chen, H.; Zhang, W.; Tang, X.-Q.; Ding, Y.-H.; Yin, J.-R.; Jiang, Y.; Zhang, P.; Jin, H. First Principles Study of P-Doped Borophene as Anode Materials for Lithium Ion Batteries. *Appl. Surf. Sci.* **2018**, *427* (Part B), 198–205.
(35) Jena, N. K.; Araujo, R. B.; Shukla, V.; Ahuja, R. Borophane as a Benchmate of Graphene: A Potential 2D Material for Anode of Li and Na-Ion Batteries. *ACS Appl. Mater. Interfaces* **2017**, *9* (19), 16148–16158.
(36) Mortazavi, B.; Rahaman, O.; Ahzi, S.; Rabczuk, T. Flat Borophene Films as Anode

Materials for Mg, Na or Li-Ion Batteries with Ultra High Capacities: A First-Principles Study. *Appl. Mater. Today* **2017**, *8*, 60–67.

(37) Izadi Vishkayi, S.; Bagheri Tagani, M. Current-Voltage Characteristics of Borophene and Borophane Sheets. *Phys. Chem. Chem. Phys.* **2017**, *19* (32), 21461–21466.

(38) Xiang, P.; Chen, X.; Zhang, W.; Li, J.; Xiao, B.; Li, L.; Deng, K. Metallic Borophene Polytypes as Lightweight Anode Materials for Non-Lithium-Ion Batteries. *Phys. Chem. Chem. Phys.* **2017**, *19* (36), 24945–24954.

(39) Zhang, L.; Liang, P.; Shu, H.; Man, X.; Li, F.; Huang, J.; Dong, Q.; Chao, D. Borophene as Efficient Sulfur Hosts for Lithium–Sulfur Batteries: Suppressing Shuttle Effect and Improving Conductivity. *J. Phys. Chem. C* **2017**, *121* (29), 15549–15555.

(40) Liu, J.; Wang, S.; Sun, Q. All-Carbon-Based Porous Topological Semimetal for Li-Ion Battery Anode Material. *Proc. Natl. Acad. Sci.* **2017**, *114* (4), 651–656.

(41) Kulish, V. V. Surface Reactivity and Vacancy Defects in Single-Layer Borophene Polymorphs. *Phys. Chem. Chem. Phys.* **2017**, *19* (18), 11273–11281.

(42) Khanifaev, J.; Pekoz, R.; Konuk, M.; Durgun, E. The Interaction of Halogen Atoms and Molecules with Borophene. *Phys. Chem. Chem. Phys.* **2017**.

(43) Penev, E. S.; Kutana, A.; Yakobson, B. I. Can Two-Dimensional Boron Superconduct? *Nano Lett.* **2016**, *16* (4), 2522–2526.

(44) Sadeghzadeh, S. Borophene Sheets with in-Plane Chain-like Boundaries; a Reactive Molecular Dynamics Study. *Comput. Mater. Sci.* **2018**, *143*, 1–14.

(45) Izadi Vishkayi, S.; Bagheri Tagani, M. Edge-Dependent Electronic and Magnetic Characteristics of Freestanding β 12-Borophene Nanoribbons. *Nano-Micro Lett.* **2017**, *10* (1), 14.

(46) Sadeghzadeh, S. The Creation of Racks and Nanopores Creation in Various Allotropes of Boron due to the Mechanical Loads. *Superlattices Microstruct.* **2017**, *111* (Supplement C), 1145–1161.

(47) Mortazavi, B.; Rahaman, O.; Dianat, A.; Rabczuk, T. Mechanical Responses of Borophene Sheets: A First-Principles Study. *Phys. Chem. Chem. Phys.* **2016**, *18* (39), 27405–27413.

(48) Adamska, L.; Sharifzadeh, S. Fine-Tuning the Optoelectronic Properties of Freestanding Borophene by Strain. *ACS Omega* **2017**, *2* (11), 8290–8299.

(49) Nishino, H.; Fujita, T.; Cuong, N. T.; Tominaka, S.; Miyauchi, M.; Iimura, S.; Hirata, A.; Umezawa, N.; Okada, S.; Nishibori, E.; Fujino, A.; Fujimori, T.; Ito, S.; Nakamura, J.; Hosono, H.; Kondo, T. Formation and Characterization of Hydrogen Boride Sheets Derived from MgB2 by Cation Exchange. *J. Am. Chem. Soc.* **2017**, *139* (39), 13761–13769.

(50) Jiao, Y.; Ma, F.; Bell, J.; Bilic, A.; Du, A. Two-Dimensional Boron Hydride Sheets: High Stability, Massless Dirac Fermions, and Excellent Mechanical Properties. *Angew. Chemie* **2016**, *128* (35), 10448–10451.

(51) Zhang, X.; Hu, J.; Cheng, Y.; Yang, H. Y.; Yao, Y.; Yang, S. A. Borophene as an Extremely High Capacity Electrode Material for Li-Ion and Na-Ion Batteries. *Nanoscale* **2016**, *8* (33), 15340–15347.

(52) Guo, Y.-G.; Hu, J.-S.; Wan, L.-J. Nanostructured Materials for Electrochemical Energy Conversion and Storage Devices. *Adv. Mater.* **2008**, *20* (15), 2878–2887.

(53) Kresse, G.; Furthmüller, J. Efficient Iterative Schemes for Ab Initio Total-Energy Calculations Using a Plane-Wave Basis Set. *Phys. Rev. B* **1996**, *54* (16), 11169–11186.

(54) Perdew, J. P.; Burke, K.; Ernzerhof, M. Generalized Gradient Approximation Made Simple. *Phys. Rev. Lett.* **1996**, *77* (18), 3865–3868.
(55) Kresse, G.; Joubert, D. From Ultrasoft Pseudopotentials to the Projector Augmented - Wave Method. *Phys. Rev. B* **1999**, *59*, 1758.
(56) Bučko, T.; Hafner, J.; Lebègue, S.; Ángyán, J. G. Improved Description of the Structure of Molecular and Layered Crystals: Ab Initio DFT Calculations with van Der Waals Corrections. *J. Phys. Chem. A* **2010**, *114* (43), 11814–11824.
(57) Wang, Z.; Selbach, S. M.; Grande, T. Van Der Waals Density Functional Study of the Energetics of Alkali Metal Intercalation in Graphite. *RSC Adv.* **2014**, *4* (8), 4069–4079.
(58) Grimme, S. Semiempirical GGA-Type Density Functional Constructed with a Long-Range Dispersion Correction. *J. Comput. Chem.* **2006**, *27* (15), 1787–1799.
(59) Tang, W.; Sanville, E.; Henkelman, G. A Grid-Based Bader Analysis Algorithm without Lattice Bias. *J. Phys. Condens. Matter* **2009**, *21* (8), 84204.
(60) Aydinol, M. K.; Kohan, A. F.; Ceder, G.; Cho, K.; Joannopoulos, J. Ab Initio Study of Lithium Intercalation in Metal Oxides and Metal Dichalcogenides. *Phys. Rev. B* **1997**, *56* (3), 1354–1365.
(61) Koudriachova, M. V.; Harrison, N. M.; De Leeuw, S. W. Diffusion of Li-Ions in Rutile. An Ab Initio Study. In *Solid State Ionics*; 2003; Vol. 157, pp 35–38.
(62) Tarascon, J. M.; Armand, M. Issues and Challenges Facing Rechargeable Lithium Batteries. *Nature* **2001**, *414* (6861), 359–367.
(63) Yang, Z.; Choi, D.; Kerisit, S.; Rosso, K. M.; Wang, D.; Zhang, J.; Graff, G.; Liu, J. Nanostructures and Lithium Electrochemical Reactivity of Lithium Titanites and Titanium Oxides: A Review. *Journal of Power Sources*. 2009, pp 588–598.
(64) Mortazavi, B.; Dianat, A.; Cuniberti, G.; Rabczuk, T. Application of Silicene, Germanene and Stanene for Na or Li Ion Storage: A Theoretical Investigation. *Electrochim. Acta* **2016**, *213*, 865–870.
(65) Zhao, S.; Kang, W.; Xue, J. The Potential Application of Phosphorene as an Anode Material in Li-Ion Batteries. *J. Mater. Chem. A* **2014**, *2* (44), 19046–19052.
(66) Mortazavi, B.; Dianat, A.; Rahaman, O.; Cuniberti, G.; Rabczuk, T. Borophene as an Anode Material for Ca, Mg, Na or Li Ion Storage: A First-Principle Study. *J. Power Sources* **2016**, *329*, 456–461.
(67) Uthaisar, C.; Barone, V. Edge Effects on the Characteristics of Li Diffusion in Graphene. *Nano Lett.* **2010**, *10* (8), 2838–2842.
(68) Makaremi, M.; Mortazavi, B.; Singh, C. V. Carbon Ene-Yne Graphyne Monolayer as an Outstanding Anode Material for Li/Na Ion Batteries. *Appl. Mater. Today* **2018**, *10*, 115–121.
(69) Tritsaris, G. A.; Kaxiras, E.; Meng, S.; Wang, E. Adsorption and Diffusion of Lithium on Layered Silicon for Li-Ion Storage. *Nano Lett.* **2013**, *13* (5), 2258–2263.
(70) Li, Y.; Wu, D.; Zhou, Z.; Cabrera, C. R.; Chen, Z. Enhanced Li Adsorption and Diffusion on MoS$_2$ Zigzag Nanoribbons by Edge Effects: A Computational Study. *J. Phys. Chem. Lett.* **2012**, *3* (16), 2221–2227.
(71) Er, D.; Li, J.; Naguib, M.; Gogotsi, Y.; Shenoy, V. B. Ti$_3$C$_2$ MXene as a High Capacity Electrode Material for Metal (Li, Na, K, Ca) Ion Batteries. *ACS Appl. Mater. Interfaces* **2014**, *6* (14), 11173–11179.
(72) Rong, Z.; Malik, R.; Canepa, P.; Sai Gautam, G.; Liu, M.; Jain, A.; Persson, K.; Ceder, G. Materials Design Rules for Multivalent Ion Mobility in Intercalation Structures. *Chem.*

*Mater.* **2015**, *27* (17), 6016–6021.
(73) Olson, C. L.; Nelson, J.; Islam, M. S. Defect Chemistry, Surface Structures, and Lithium Insertion in Anatase TiO2. *J. Phys. Chem. B* **2006**, *110* (20), 9995–10001.
(74) Persson, K.; Hinuma, Y.; Meng, Y. S.; Van der Ven, A.; Ceder, G. Thermodynamic and Kinetic Properties of the Li-Graphite System from First-Principles Calculations. *Phys. Rev. B* **2010**, *82* (12), 125416.

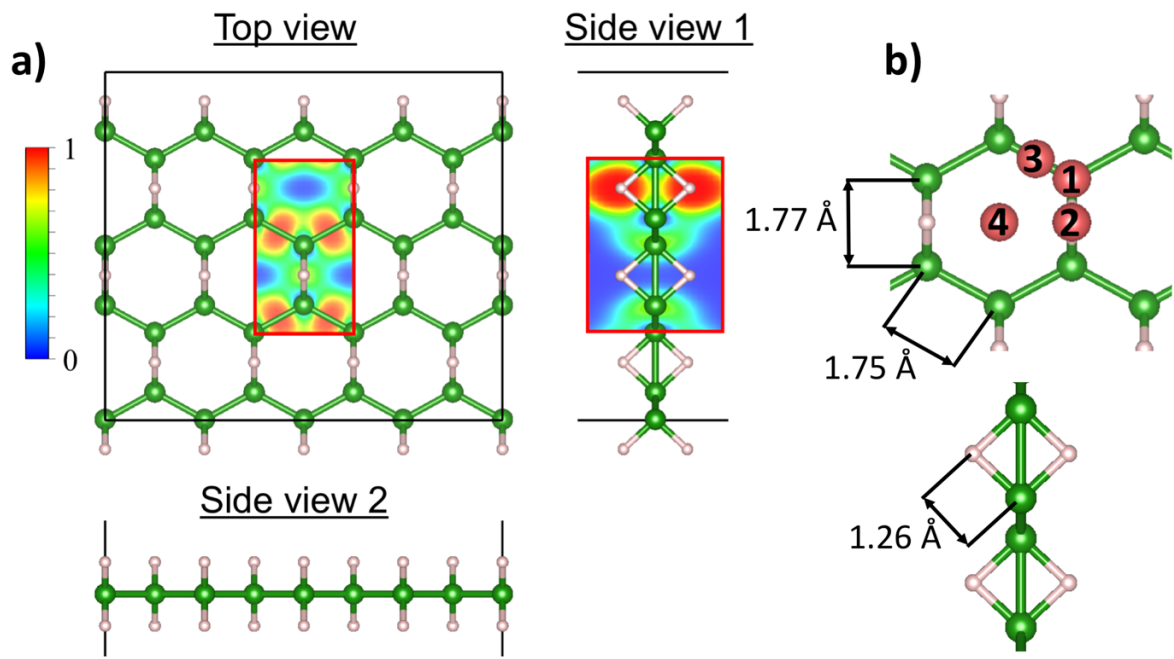

Figure 1. a) Hydrogen boride (HB) structural configuration including the supercell (black line) and the unitcell (red line) with lattice constants a = 3.02 Å and b = 5.29 Å. The contours illustrate electron localization function (ELF), which has a value between 0 and 1, where 1 corresponds to the perfect localization. b) Different possible adatom adsorption sites on the HB surface. Green, white, and red balls represent boron, hydrogen, and adatom, respectively.

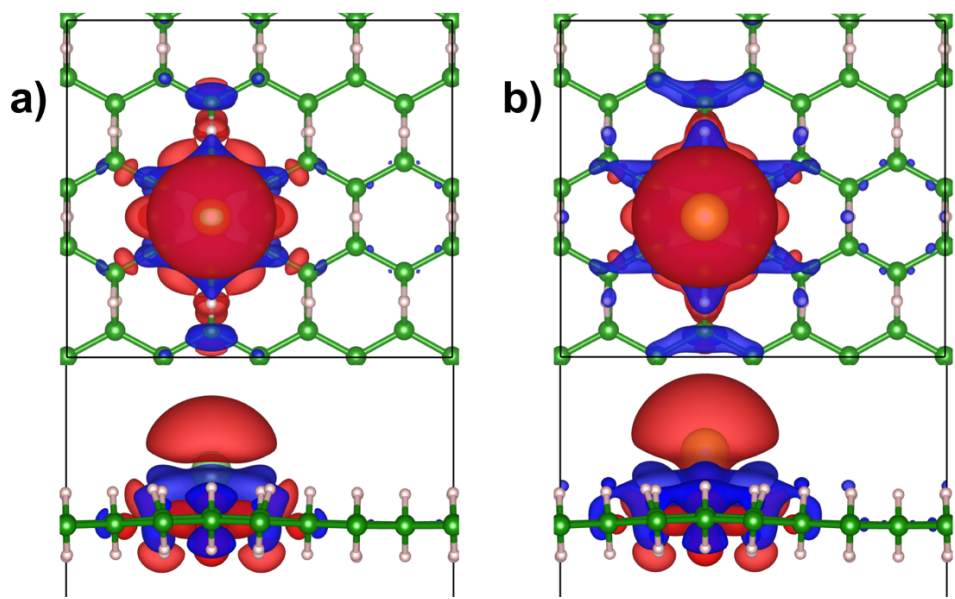

Figure 2. Binding charge transfer due to the adsorption of a) Li and b) Na to the hexagonal site of the monolayer. Color coding involves blue for charge sufficient and red for the charge deficient regions; respectively, (isosurface value is 0.001 e/Å$^3$).

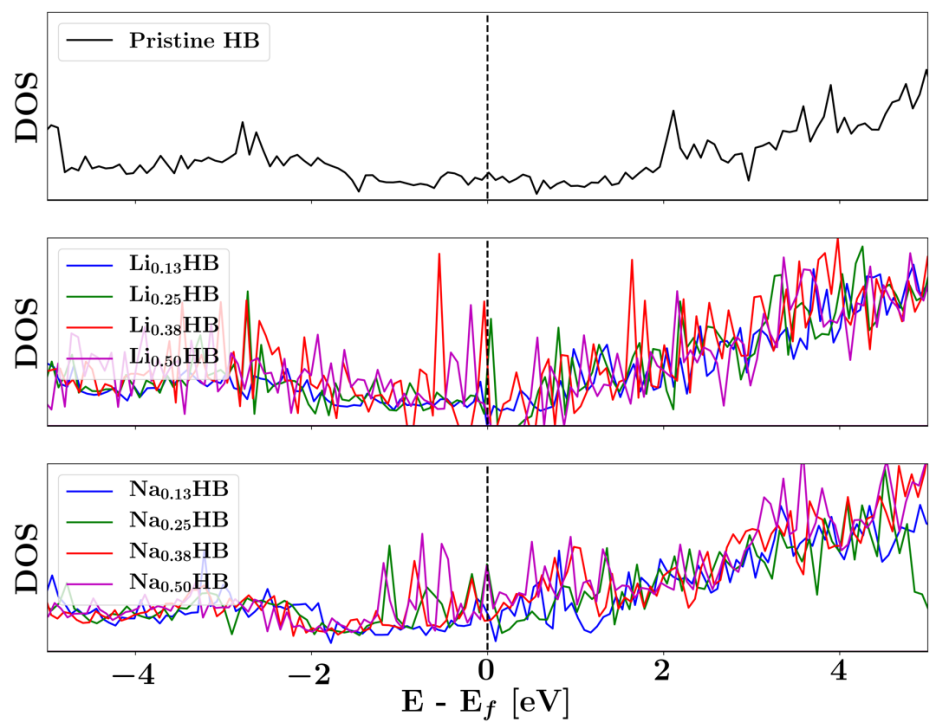

Figure 3. Density of states (DOS) for the pristine HB monolayer, and monolayers interacting with different amount of adatom inserion. The black dashed line represents the Fermi-level.

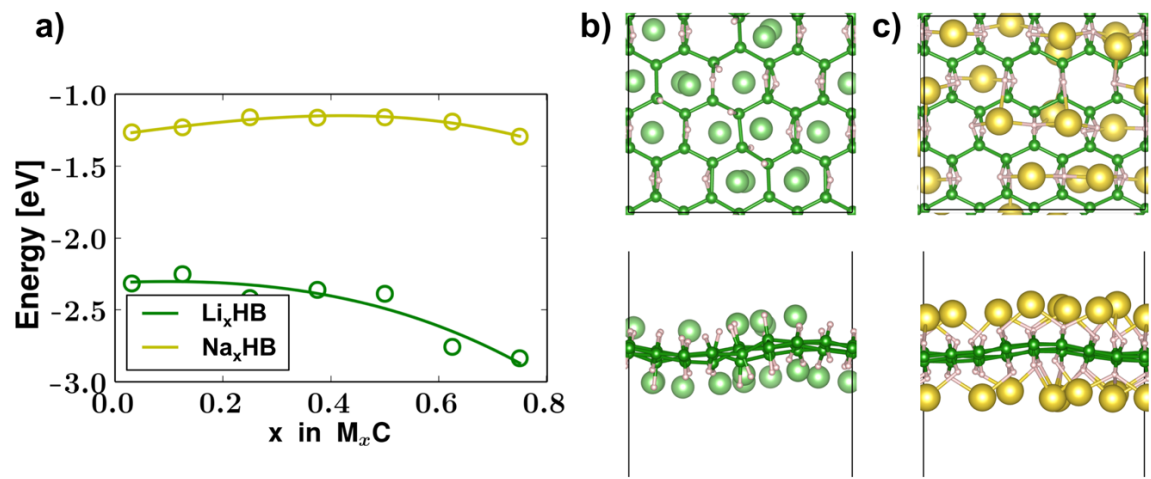

Figure 4. Adsorption of adatoms by the HB surface. a) Binding energy of adatoms with respect to the coverage ratio of x. The adsorption of b) Li and c) Na adatoms to the anode surface at x < 0.5. Color coding involves green and yellow for Li and Na, respectively.

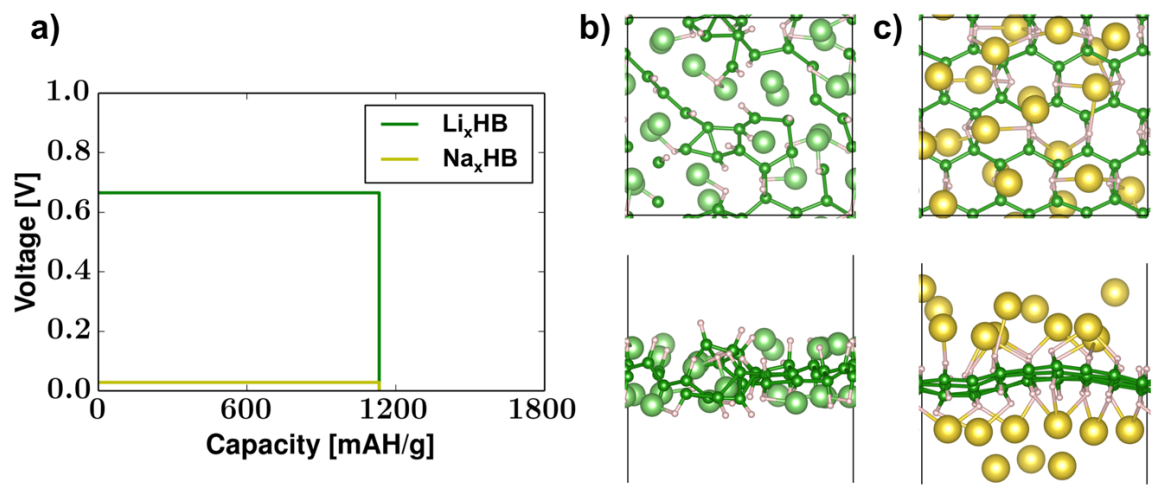

Figure 5. Storage capacity of the HB nanosheet. a) Average open-circuit voltage the anode with respect to the capacity. The nanosheet-adatoms configuration after reaching the optimal capacity (x > 0.5) due to the insertion of b) Li and c) Na adatoms. Color coding is identical to Figure 4.

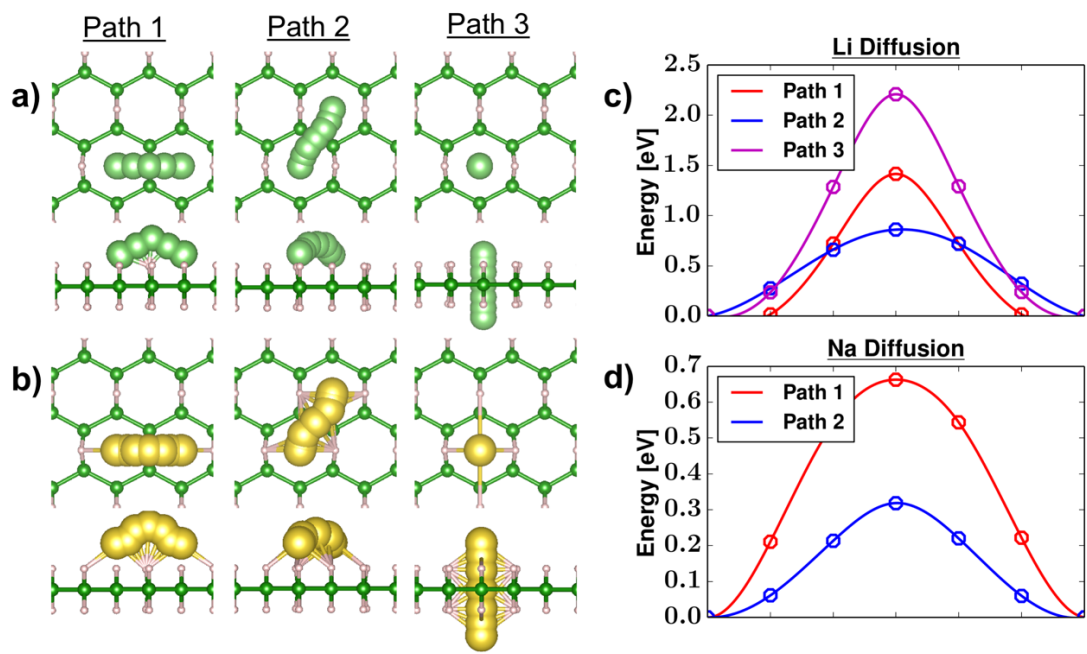

Figure 6. Intercalation of single adatoms into the HB nanosheet through different pathways (Path 1, Path 2, and Path 3). a) Li diffusion and b) Na diffusion; Nudged-elastic band (NEB) energy curvatures for c) Li and d) Na adatoms. Color coding of atoms is similar to Figure 4.

# Supporting Information

# 2D Hydrogenated Graphene-like Borophene as a High Capacity Anode Material for Improved Li/Na Ion Batteries: A First Principles Study

Meysam Makaremi,[1] Bohayra Mortazavi,[2] and Chandra Veer Singh*[,1,3]

[1]Department of Materials Science and Engineering, University of Toronto, 184 College Street, Suite 140, Toronto, ON M5S 3E4, Canada.
[2]Institute of Structural Mechanics, Bauhaus-Universität Weimar, Marienstr. 15, D-99423 Weimar, Germany.
[3]Department of Mechanical and Industrial Engineering, University of Toronto, 5 King's College Road, Toronto M5S 3G8, Canada.

1) **Hydrogen Boride Structural Details**

Table S1: Lattice vectors of the hydrogen boride unitcell.

| Vector | x [Å] | y [Å] | z [Å] |
|---|---|---|---|
| 1 | 3.017 | 0.000 | 0.000 |
| 2 | 0.000 | 5.291 | 0.000 |
| 3 | 0.000 | 0.000 | 20.000 |

Table S2: Atomic positions of the hydrogen boride unitcell.

| Atom | x [Å] | y [Å] | z [Å] |
|---|---|---|---|
| B | 0.000 | 0.000 | 10.000 |
| B | 1.508 | 0.881 | 10.000 |
| B | 1.508 | 2.645 | 10.000 |
| B | 0.000 | 3.527 | 10.000 |
| H | 0.000 | 4.409 | 9.090 |
| H | 1.508 | 1.763 | 10.900 |
| H | 1.508 | 1.763 | 9.090 |
| H | 0.000 | 4.409 | 10.900 |

2) <u>**Li Adsorption on Hydrogen Boride**</u>

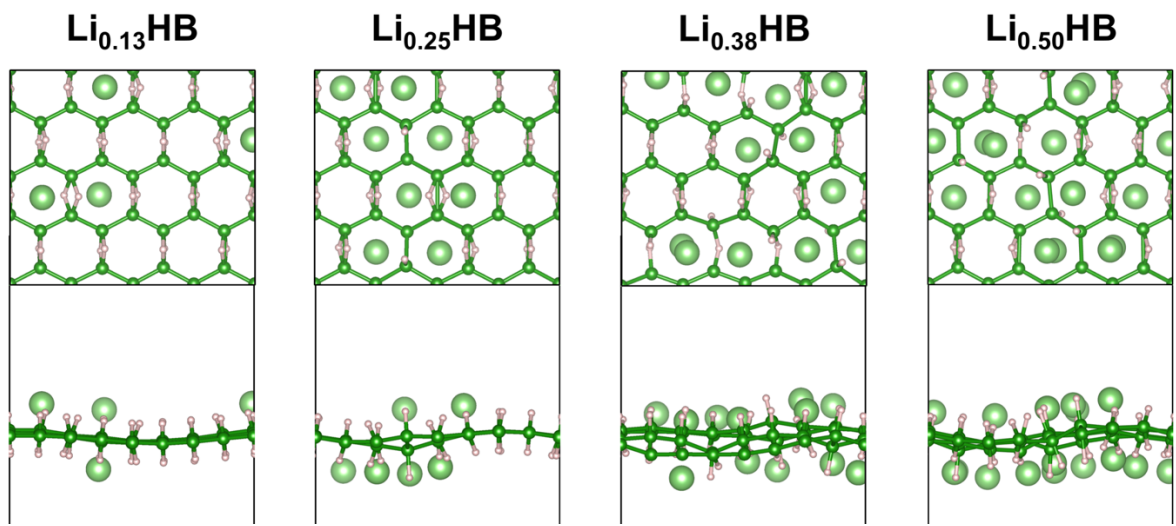

Figure S1. Adsorption of Li adatoms on HB nanosheet. Color coding is identical to Figure 4.

## 3) Na Adsorption on Hydrogen Boride

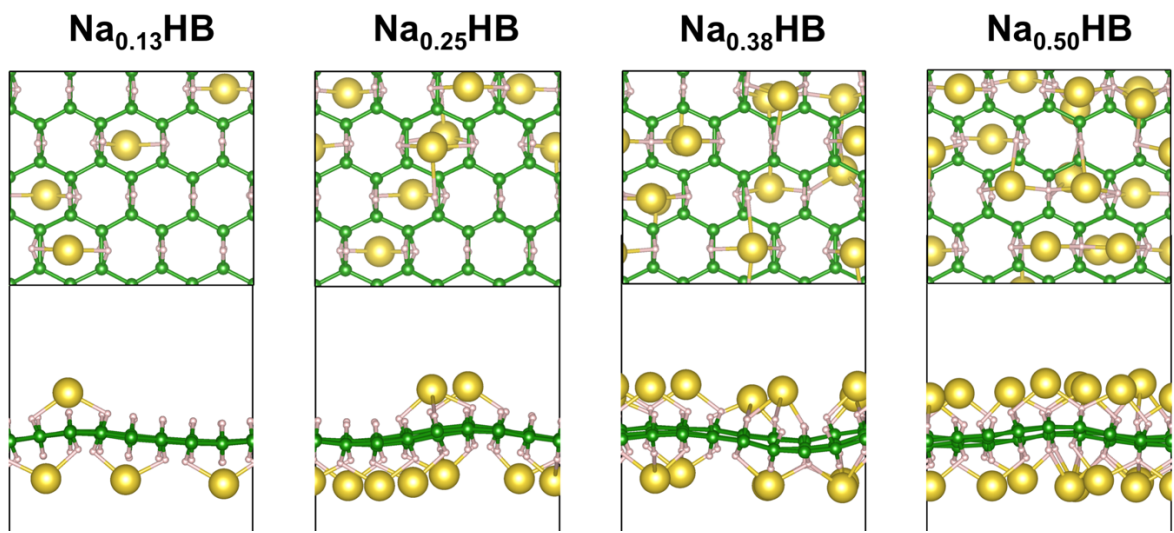

Figure S2. Adsorption of Na adatoms on HB nanosheet. Color coding is identical to Figure 4.

4) **Electronic Charge Density of Before and After Optimal Storage Capacity for Na Insertion**

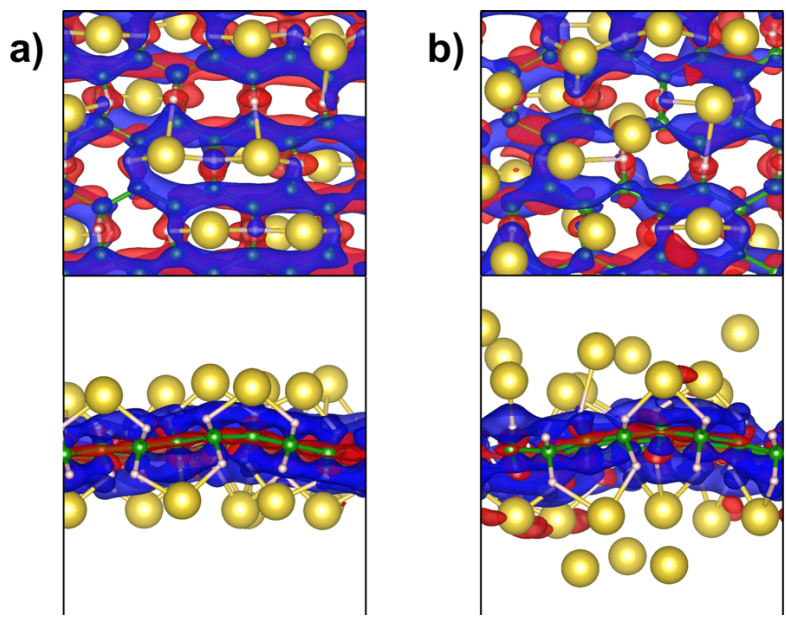

Figure S3. Differential Charge density of the HB nanosheet interacting with Na adatoms. The charge difference configuration a) before and b) after reaching the optimal capacity due to the insertion of Na adatoms. Color coding is identical to Figures 2 and 4.

## 5) **Thermal Stability Analysis by AIMD Simulations**

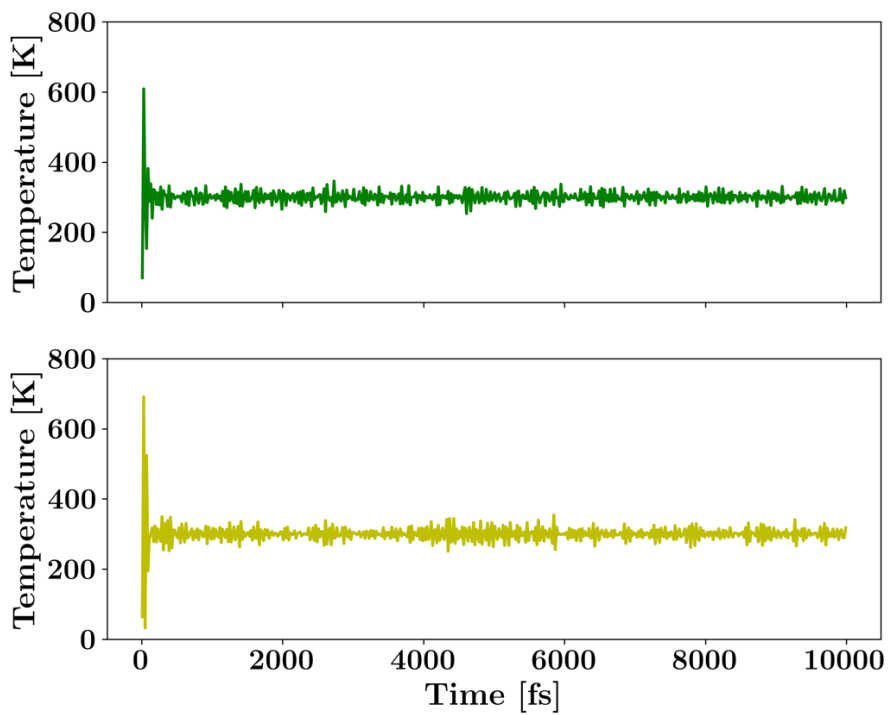

Figure S4. Thermal stability analysis. The variation of the temperature with respect to the simulation time for the HB monolayer interacting with Li (green)/Na (yellow) adatoms at the optimal capacity.